\documentclass[useAMS,usenatbib,usegraphicx]{mn2e}

\usepackage{times}

\newcommand{\smallcolsize}{0.40}
\newcommand{\singlecolsize}{0.47}
\newcommand{\middlecolsize}{0.80}
\newcommand{\doublecolsize}{0.90}

\newcommand{\Vmax}{V_{\rm max}}
\newcommand{\D}{{\rm d}}
\newcommand{\mass}{{\cal M}}
\newcommand{\Msun}{{\cal M}_{\odot}}
\newcommand{\Lsun}{L_{\odot}}

\newcommand{\hunits}{{\rm\,km\,s^{-1}\,Mpc^{-1}}}

\newcommand{\araa}{ARA\&A}   \newcommand{\aap}{A\&A}
\newcommand{\aj}{AJ}         \newcommand{\apj}{ApJ}
\newcommand{\apjl}{ApJ}      \newcommand{\apjs}{ApJS}
\newcommand{\mnras}{MNRAS}   \newcommand{\nat}{Nature}
     \newcommand{\pasp}{PASP}

\newcommand{\pegase}{{\small PEGASE}}
\newcommand{\basti}{{\small BaSTI}}
\newcommand{\HI}{H{\small I}}
\newcommand{\HIcaption}{H{\scriptsize I}}

\title[Galaxy mass functions]
{On the galaxy stellar mass function, 
 the mass-metallicity relation, 
 and the implied baryonic mass function}

\author[I.~K.~Baldry et al.]
{I.~K.~Baldry$^1$, K.~Glazebrook$^2$, S.~P.~Driver$^3$ \\
$^1$Astrophysics Research Institute, Liverpool John Moores University, 
  Twelve Quays House, Egerton Wharf, Birkenhead CH41~1LD, UK\\
$^2$Centre for Astrophysics and Supercomputing, 
   Swinburne University of Technology, 
   Mail~31, P.O.\ Box~218, Hawthorne, VIC~3122, Australia\\
$^3$SUPA, School of Physics and Astronomy, University of St Andrews, 
  North Haugh, St Andrews, Fife, KY16~9SS, UK}

\begin{document}

\date{Accepted by MNRAS 2008 April 16. 
Received 2008 April 08; in original form 2008 February 07}

\pagerange{\pageref{firstpage}--\pageref{lastpage}} \pubyear{2008}

\maketitle

\label{firstpage}

\begin{abstract}
  A comparison between published field galaxy stellar mass functions (GSMFs)
  shows that the cosmic stellar mass density is in the range 4--8 per cent of
  the baryon density (assuming $\Omega_b=0.045$).  There remain significant
  sources of uncertainty for the dust correction and underlying stellar
  mass-to-light ratio even assuming a reasonable universal stellar initial
  mass function.  We determine the $z<0.05$ GSMF using the New York University
  -- Value-Added Galaxy Catalog sample of 49968 galaxies derived from the
  Sloan Digital Sky Survey and various estimates of stellar mass. The GSMF
  shows clear evidence for a low-mass upturn and is fitted with a double
  Schechter function that has $\alpha_2 \simeq -1.6$.  At masses below $\sim
  10^{8.5}\Msun$, the GSMF may be significantly incomplete because of missing
  low surface-brightness galaxies.  One interpretation of the stellar
  mass-metallicity relation is that it is primarily caused by a lower fraction
  of available baryons converted to stars in low-mass galaxies. Using this
  principal, we determine a simple relationship between baryonic mass and
  stellar mass and present an `implied baryonic mass function'. This function
  has a faint-end slope, $\alpha_2 \simeq -1.9$.  Thus, we find evidence that
  the slope of the low-mass end of the galaxy mass function could plausibly be
  as steep as the halo mass function. We illustrate the relationship between
  halo baryonic mass function $\rightarrow$ galaxy baryonic mass function
  $\rightarrow$ GSMF. This demonstrates the requirement for peak galaxy
  formation efficiency at baryonic masses $\sim 10^{11}\Msun$ corresponding to
  a minimum in feedback effects. The baryonic-infall efficiency may have
  levelled off at lower masses.
 \end{abstract}


\section{Introduction}
\label{sec:intro}

The galaxy luminosity or mass function is a fundamental tool used in
interpreting the evolution of galaxies. The functions are usually
defined as the number density of galaxies per logarithmic luminosity
or mass interval. A steeply rising mass function to the faint
population has been a generic prediction of galaxy formation based on
hierarchical clustering \citep{WR78,KWG93,cole94}.  In contrast, the
field galaxy luminosity function was observed to have a significantly
flatter `faint end' \citep{BST88,loveday92}. In order to reconcile
cold dark matter (CDM) galaxy formation models with the observed
luminosity function, star formation (SF) is suppressed in low-mass halos
by, for example, supernovae feedback \citep{LS91,kay02} or
photoionisation \citep{Efstathiou92,Somerville02}.

On the observational side, accurately determining the number densities of
faint galaxies in the field or in clusters is challenging. This is mainly
because of the low surface brightness (SB) of these galaxies, which means that
they can be undetected in photometry even if high-SB galaxies with the same
apparent magnitude are detected \citep{Disney76,DP83}. Despite these and other
challenges, there is evidence for a `faint-end upturn' in luminosity functions
whereby the luminosity function is rising steeply fainter than about 3--5
magnitudes below the characteristic luminosity, in clusters
\citep{driver94,popesso05} or the field
\citep{Loveday97,zucca97,marzke98,blanton05}.  Note however that the upturn is
not always evident \citep{norberg02,HdP07}.

A faint-end upturn suggests that the efficiency of feedback has levelled off
at low galaxy luminosities. A more direct analysis is to compare galaxy mass
functions with predicted mass functions from CDM models. \citet{klypin99}
compared the circular velocity distribution of satellite galaxies in the Local
Group (see also \citealt{moore99substructure}). This highlights the
`substructure problem' where there are 5--10 times as many low-mass satellites
in CDM models than observed. However, we still have not reached a complete
census of Local Group satellites as evidenced by recent discoveries of low-SB
galaxies around M31 \citep{martin06} and the Milky Way \citep{belokurov07}.
 
Reliable dynamical mass estimates for complete and large samples of
field galaxies are difficult to obtain. Stellar masses can be
estimated for significantly larger samples of galaxies using the
principles of stellar population synthesis \citep{TG76,Tinsley80}.
Thus, galaxy stellar mass functions (GSMFs) can be derived from galaxy
luminosities (e.g., \citealt{balogh01,cole01,bell03,KB03}).
Comparisons between GSMFs are then, in theory, free of the
stellar-population contribution to mass-to-light (M-L) variations that
are inherent in comparisons between luminosity functions.

Integrating a field GSMF to determine the cosmic stellar mass density (SMD)
brings to light the large gap between this value ($\Omega_{\rm stars} \sim
0.003$) and estimates of $\Omega_b$ from Big-Bang Nucleosynthesis theory
(0.04, \citealt*{BNT01}) or the power spectrum of the cosmic background
radiation (0.045, \citealt{spergel07}).  Even determination of the baryonic
content of galaxies including stars and cold gas accounts for less than or
about 0.1 of $\Omega_b$ \citep{bell03gbmf}.  About 0.2 of $\Omega_b$ is
accounted for by hot plasma identified by X-ray emission in clusters and
groups \citep*{FHP98}; while the rest is projected to be in a more diffuse
inter-galactic medium \citep{CO99} and perhaps the `coronae' of galaxies
\citep{MB04}.  The overall efficiency of baryons falling into the luminous
disks or bulges of galaxies is low.

The efficiency of SF (fraction of baryonic mass converted to stars) varies
significantly with galaxy mass.  A low efficiency averaged over the life of a
galaxy gives rise to a low gas-phase metallicity (in the inter-stellar medium)
because the metal production by supernovae is diluted by gas reservoirs within
a galaxy \citep{Tinsley80,brooks07} and further infalling material. Conversely
a high efficiency can drive the gas-phase metallicity to high values.  In this
paper, we explore the use of the stellar mass-metallicity relation
\citep{tremonti04} as a SF efficiency estimator to convert the GSMF to a
baryonic mass function.

The plan of this paper follows.  Determinations of the field GSMF and the SMD
are reviewed in \S~\ref{sec:compare-gsmf}.  An estimate of the low-redshift
field GSMF paying careful attention to SB selection effects is described in
\S~\ref{sec:nyu-vagc}.  The evident faint-end upturn is compared with cluster
GSMFs. The relationship between gas-phase metallicity and stellar mass is then
used to convert the field GSMF to an `implied baryonic mass function' assuming
a simple relationship between metallicity and the fraction of baryonic mass in
stars. This is described in \S~\ref{sec:mzr}. This galaxy baryonic mass
function, and the GSMF, are compared with halo mass functions and related to
galaxy formation efficiency as a function of mass.  Summary and conclusions
are presented in \S~\ref{sec:conclusions}.  The dependencies of stellar M-L
ratio on various assumptions are presented in the Appendix.  Throughout this
paper, a cosmology with $H_0=70\hunits$, $\Omega_{m,0}=0.3$ and
$\Omega_{\Lambda,0}=0.7$ is assumed.

\section{Comparison between published galaxy stellar mass functions}
\label{sec:compare-gsmf}

By estimating stellar M-L ratios for galaxies it is possible to calculate the
equivalent of a luminosity function, for the total stellar mass of galaxies,
known as the galaxy stellar mass function \citep{cole01}. This then gives a
more fundamental account of the baryons that are locked up in stars and how
they are distributed amongst galaxies of different masses.

One of the key ingredients in this calculation is the stellar initial mass
function (IMF), which is generally assumed to be independent of galaxy type or
mass \citep{Elmegreen01}. For the comparisons in this paper, the IMFs used are
the `diet Salpeter' that is defined as $0.7\,\times$ the mass derived from a
standard Salpeter \citep{BdJ01,bell03}, the \citet{Kroupa01}, and the
\citet{Chabrier03} IMF.  These are all similar in terms of M-L ratio as a
function of galaxy colour. The variations highlighted and discussed between
different mass estimates in this paper are not significantly dependent on IMF
choice (Appendix).

\begin{figure}
  \includegraphics[width=\singlecolsize\textwidth]{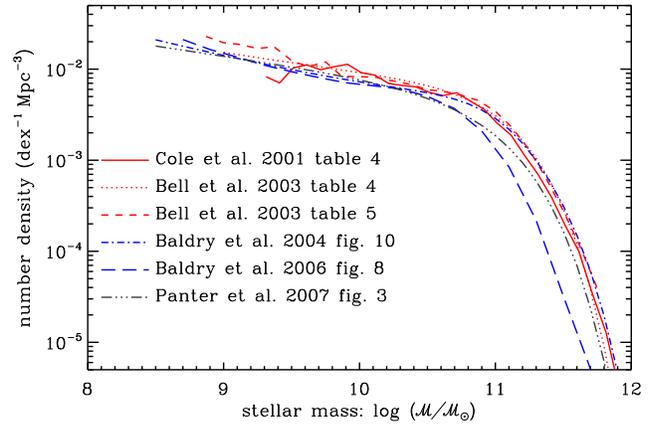}
\caption{Comparison between published field galaxy stellar mass
  functions ($z\la0.1$).  The mass functions have been converted to
  $H_0=70\hunits$ where necessary and the \citeauthor{cole01}\ GSMF has been
  converted to the diet Salpeter IMF.}
\label{fig:compare-gsmf}
\end{figure}

Figure~\ref{fig:compare-gsmf} shows a comparison between published
field GSMFs (where `field' in this case means a cosmic volume
average). A brief description of how these were derived is given
below. Masses and number densities were converted to a cosmology with
$H_0=70\hunits$ where necessary.

\begin{enumerate}
\item \citet{cole01}: data were taken from the Salpeter column of
  table~4. The masses were multiplied by 0.7 to convert to the diet
  Salpeter IMF. The published GSMF was derived from a match of the Two
  Micron All Sky Survey (2MASS) extended source catalogue
  \citep{jarrett00} to the 2dF Galaxy Redshift Survey (2dFGRS;
  \citealt{colless01}).  \citeauthor{cole01} computed $b_J-K$ and
  $J-K$ colours for a range of exponentially declining 
  SF histories with $z_{\rm form}=20$ and metallicities using
  \citet{BC93} models.  Dust attenuation from the models of
  \citet{ferrara99} were applied.  Stellar M-L ratios in the near-IR
  bands were determined from the models that most closely matched the
  observed galaxies' colours; with stellar masses assumed to be 0.72
  times the integral of the SF rate (to account for recycling).
%% M/L_K implied to be 1.32, see p. 271 of Cole et al. 2001
\item \citet{bell03}: data were taken from table~4.  The plotted GSMF
  is the sum of the $g$-band derived early- and late-type Schechter
  functions. The GSMFs were derived from a match of the 2MASS
  catalogue to the Sloan Digital Sky Survey (SDSS;
  \citealt{york00,stoughton02}). The $ugrizK$ data were fitted to
  magnitudes computed for a range of exponential SF histories and
  metallicities using \pegase\ \citep{FR97,FR99} models.  M-L ratios
  were determined from the best-fit models.
\item \citeauthor{bell03}: as above but using data taken from table~5. 
  This is a binned GSMF derived from the same M-L ratios.
\item \citet{baldry04}: data were taken from fig.~10.  The GSMF is the
  sum of the red- and blue-sequence Schechter functions.  This was
  derived from SDSS data with a M-L ratio given by a linear function
  of colour: $\log (\mass/{L_r}) \:=\: -0.55 \, + \, 0.45 \, (u-r)
  \,$.\footnote{$\mass$ or $\mass_s$ 
    is used for stellar mass in this paper.} This
  relation was determined from stellar masses computed by
  \citet{bell03} and \citet{kauffmann03A}.
\item \citet{baldry06}: data were taken from fig.~8. The GSMF is the
  sum over all the environments. This was derived in a similar way to
  that of \citet{baldry04} using a different relationship between M-L ratio
  and colour. In this case, the relation was determined from stellar
  masses computed by \citet{kauffmann03A} and K.G.  These methods are
  described in \S~\ref{sec:stellar-masses}.
\item \citet{panter07}: data were taken from fig.~3. The GSMF is
  derived from the Schechter parameters. The method for estimating
  stellar masses is described in \S~\ref{sec:stellar-masses}.
\end{enumerate}

The GSMFs are similar except for the \citet{baldry06} function, which
appears to be shifted by about 0.2\,dex to lower masses. This is
because the method used by K.G.\ for that paper underestimated
the masses of luminous red galaxies. This may reflect degeneracies
with fitting to only $ugriz$ photometry and attempting to fit SF
bursts, metallicity and dust attenuation (Appendix).

\subsection{The stellar mass density of the Universe}
\label{sec:smd}

A quantitative overall comparison can be made by integrating each GSMF
to obtain the total SMD.  The critical density for $H_0=70\hunits$ is $\rho_c
= 1.36 \times 10^{11} \Msun {\rm Mpc}^{-3}$ and we assume $\Omega_{b}
= 0.045$ \citep{spergel07}. The fraction of baryons in stars $f_s =
\Omega_{\rm stars} / \Omega_{b}$ is then (i) 0.054, (ii) 0.061, (iii)
0.063, (iv) 0.056, (v), 0.035, (vi) 0.042. Note these values do {\em
not} include extrapolations of the GSMF to lower masses than the
plotted points or corrections to total galaxy magnitudes from
catalogue magnitudes unless already applied.\footnote{SDSS Petrosian
magnitudes theoretically recover 99\% of the flux of a galaxy with an
exponential profile, 80\% with a de-Vaucouleurs profile, and 95\%
in the case of nearly unresolved systems (close to the point-spread
function of SDSS imaging) \citep{blanton01,stoughton02}.
\citet{bell03} have used estimated corrections for this loss of
light.\label{ftn:petrosian}}
%% SMD values with extrapolation to faint-end
%% (i) 0.057, (ii) 0.062, 0.064, (iii) 0.056, (iv) 0.035, (v) 0.043

The total SMD can also be determined using methods other than
integrating a GSMF.  \citet{driver07} determined $f_s = 0.083\pm0.012$
from the $B$-band luminosity function using an empirical
geometrical-based correction for dust attenuation; \citet{BG03} give a
range from 0.035 to 0.063 from fitting to the local cosmic luminosity
densities (over a range of IMFs but similar to those used here in
terms of M-L ratio); and \citet{nagamine06} found $0.051\pm0.009$ from
fitting to various data including the extragalactic background light.
See also table~5 of \citet{gallazzi08} for a list of SMD measurements.

Overall there remains a considerable uncertainty in $\Omega_{\rm
stars}$ or $f_s$, which is probably in the range 4--8 per
cent (for the Chabrier, Kroupa or diet Salpeter IMF). 
This is much larger than the uncertainty in $\Omega_m$ or
$\Omega_b$ within the standard cosmological paradigm. 

Accounting for dust is obviously a major problem.  Correcting for
attenuation by incorporating a dust law in stellar population
synthesis (PS) models, and fitting to colours and/or spectral
features, cannot account for populations behind opaque screens.
Notably, \citet{driver07} obtain the largest value of $f_s$ and the
method uses comparisons of disks and bulges viewed at various
orientations to constrain their dust model \citep{tuffs04}. It is then
useful to compare the SMD with near-IR measurements that are minimally
affected by dust attenuation.  The total $K$-band luminosity density,
in units of $10^{8} \Lsun {\rm Mpc}^{-3}$, is given by 4.0
\citep{cole01}, 5.0 \citep{kochanek01} and 4.1
\citep{bell03}.\footnote{\citet{huang03} reported a high value for the
$K$-band luminosity density of $\sim8.0$ ($10^{8} \Lsun {\rm
Mpc}^{-3}$). However a reanalysis by K.G., of the Hawaii-AAO data
extended to $K<16$, optimised for the accuracy of the luminosity
density at $z<0.2$ gives a value of $\sim4.5$ in agreement with the
2MASS results.}  Using these values and the full range of
$f_s=0.035$--0.083 implies a cosmic $\log (\mass/{L_K})$ in the range
$-0.35$ to $0.1$.  A high SMD with $f_s\sim0.08$ appears unlikely
given that only the oldest simple stellar populations have $\log
(\mass/{L_K}) \simeq 0.0$ (Appendix); unless there is more than
nominal attenuation of the $K$-band and/or the $K$-band luminosity
density is underestimated.

How can we reconcile the \citet{driver07} result ($0.083\pm0.012$)
with the standard PS methods of Fig.~\ref{fig:compare-gsmf}
(0.04--0.06 ignoring extremes) other than resorting to saying the
higher result is 2-sigma too high?  The standard PS methods cannot
account for populations behind optically-thick screens and use less
deep photometry than the \citeauthor{driver07}\ result, which is based
on the Millennium Galaxy Catalogue (MGC). A reasonable estimate of
unaccounted for light would be 20\% bringing $f_s$ up to 0.05--0.07
for the standard PS methods.\footnote{The estimated corrections to the
SMD values derived from the GSMFs of Fig.~\ref{fig:compare-gsmf} are
about 5--10\% to account for missing low-SB galaxies, low-mass
galaxies and/or corrections to total magnitudes, and 10--15\% for
optically-thick regions. The latter corresponds to approximately the
attenuation in the $K$-band derived from \citet{driver07}.}  On the
other hand, the MGC estimate is subject to larger cosmic variance and
the luminosity density appears to be high by 10--20\%,\footnote{The
uncorrected total $B$-band luminosity densities are 1.6 from MGC
\citep{driver07} and 1.3 from 2dFGRS \citep{norberg02} in units of
$10^{8} \Lsun {\rm Mpc}^{-3}$ ($H_0=70\hunits$).} and the M-L ratios
assumed for the attenuation-corrected magnitudes could also be high by
10\%.  Accounting for these brings $f_s$ down to 0.06--0.07 in
agreement with the standard PS methods.

While the largest contribution to the SMD comes from galaxies around
the break in the GSMF, lower mass galaxies play a key role in the
processing of baryons. In the next section, we discuss a new calculation
of GSMFs using various stellar mass estimates and consider how
accurately the lower-mass end can be determined.

\section{Galaxy stellar mass functions 
  from the NYU-VAGC low redshift galaxy sample}
\label{sec:nyu-vagc}

A large low-redshift sample derived from the SDSS is the New York
University -- Value-Added Galaxy Catalog (NYU-VAGC)
\citep{blanton05,blanton05nyuvagc}.\footnote{NYU-VAGC data are
available from \\ http://sdss.physics.nyu.edu/vagc/} While the data are
obtained from standard SDSS catalogues, the images have been carefully
checked for artifacts including deblending.  The data cover
cosmological redshifts from 0.0033 to 0.05 where the redshifts have
been corrected for peculiar velocities using a local Hubble-flow model
\citep{willick97}. We use the NYU-VAGC low-$z$ galaxy sample to
recompute the GSMF down to low masses.

The Data Release Four (DR4) version of the NYU-VAGC low-$z$ sample
includes data for 49968 galaxies. These data were matched to stellar
masses estimated by \citet{kauffmann03A}, \citet{gallazzi05}, and
\citet{panter07}; with 49473, 32473 and 38526 matches, respectively.
Minor adjustments were determined to account for variations between
the SDSS data from which the stellar masses were determined and the
NYU-VAGC (photometry and cosmological redshifts). Where no stellar
mass was available for a galaxy, the stellar mass was determined using
a colour-M-L relation calibrated to the particular set of stellar
masses. In addition, stellar masses were computed by K.G.\ using the
NYU-VAGC Petrosian magnitudes.  Thus there are four stellar masses for
each galaxy. The methods are described briefly below.
\label{sec:stellar-masses}

\begin{enumerate}
\item Kauffmann: The stellar masses were obtained by
fitting a grid of population synthesis models, including bursts, to
the spectral features D4000 and H$\delta$ absorption.  The predicted
colours were then compared with broad-band photometry to estimate dust
attenuation. Stellar M-L ratios were determined and applied to the
Petrosian $z$-band magnitude. For details see
\citet{kauffmann03A}.\footnote{Stellar mass estimates from
\citeauthor{kauffmann03A}\ are available from \\
http://www.mpa-garching.mpg.de/SDSS/DR4/
\label{ftn:kauffmann-masses}}
\item Gallazzi: The computation was similar to the above method except
  five spectral features were used. The features were carefully chosen
  for calculating stellar metallicity and light-weighted ages while
  minimising effects caused by chemical abundance ratios.  For details
  see \citet{gallazzi05}.\footnote{We used the DR4 catalogue of
  \citeauthor{gallazzi05}\ with stellar masses derived from $z$-band
  Petrosian magnitudes. See footnote~\ref{ftn:kauffmann-masses}
  for website.}
\item Panter: The computation involves fitting synthetic stellar
  populations to each galaxy spectrum using the {\small MOPED} data
  compression technique \citep*{HJL00}. The SF history of each galaxy
  was modelled using 11 logarithmically-spaced bins in time, each with
  an SF rate and metallicity, and a simple dust screen.  The total
  stellar mass is that obtained from the \citet{BC03} models given the
  best-fitting SF history as input. For details see
  \citet{PHJ04,panter07} and references therein.
\item Glazebrook: This is the only purely photometric method.  Stellar
  masses were determined by fitting to the observed-frame Petrosian
  $ugriz$ magnitudes of each galaxy. A grid of colours were computed
  using \pegase\ models.  A range of exponential SF histories and
  metallicities were input. Bursts were added with mass ranging from
  $10^{-4}$ to twice the mass of the primary component
  \citep{glazebrook04}.  For this paper, no dust attenuation was
  included in the models (see Appendix for discussion). Depending on
  the attenuation law, incorporating dust can be close to neutral in
  terms of M-L ratio versus colour (\citealt{BdJ01}; see fig.~12 of
  \citealt{driver07} for the effect modelled from face-on to edge-on).
\end{enumerate}

The IMFs used were the similar Kroupa and Chabrier IMFs, 
for methods (i,iv) and methods (ii,iii), respectively.
Comparing the different mass estimates for each galaxy, the standard
deviation is typically in the range 0.05--0.15 dex.  The standard
deviation is generally lower for red-sequence galaxies compared to
blue-sequence galaxies.

\begin{figure}
  \includegraphics[width=\singlecolsize\textwidth]{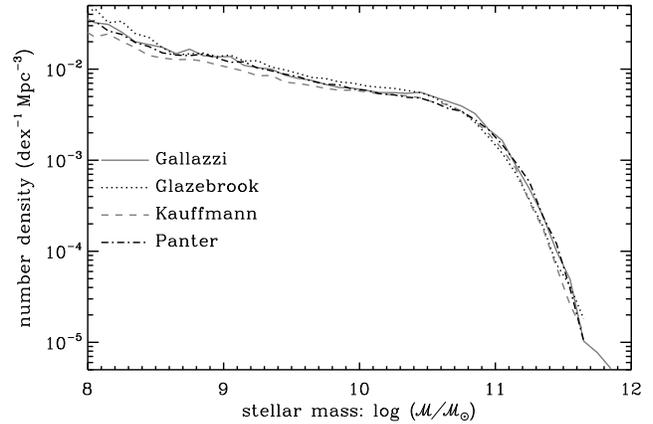}
\caption{Comparison between galaxy stellar mass functions derived from
  the NYU-VAGC low-$z$ galaxy sample.  The lines represent different
  methods of determining M-L ratios from spectra or colours. The
  functions were determined by summing $1/\Vmax$ in logarithmic mass
  bins.}
\label{fig:compare-gsmf-new}
\end{figure}

For each set of stellar masses, the galaxies were divided 
into logarithmic stellar mass bins. For each bin, the GSMF is 
then given by
\begin{equation}
 \phi_{\log \mass}  = \frac{1}{\Delta \log \mass} \,
  \sum_i \frac{1}{V_{{\rm max},i}} w_i
\label{eqn:gsmf-binned}
\end{equation}
where: $V_{{\rm max},i}$ is the comoving volume over which the $i$th
galaxy could be observed, and $w_i$ is any weight applied to the
galaxy.  The $\Vmax$ values were obtained from the NYU-VAGC catalogue
(\S~5 of
\citealt{blanton05nyuvagc}). Figure~\ref{fig:compare-gsmf-new} shows
the GSMFs derived from the NYU-VAGC using the different stellar mass
estimates. For these, no weighting was applied to any galaxy ($w=1$).

\begin{figure}
  \includegraphics[width=\singlecolsize\textwidth]{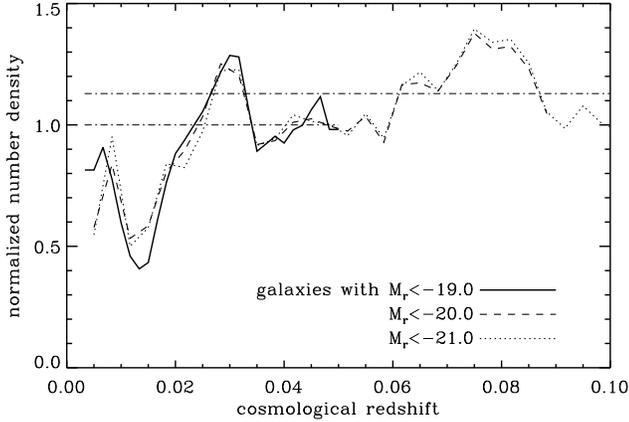}
\caption{LSS fluctuations in the low-$z$ galaxy sample.  The solid
  line shows the relative number density of galaxies brighter than
  $M_r=-19$ in redshift shells.  The number densities were normalised
  by the volume average over $z<0.05$.  The dashed and dotted lines
  show the normalised densities for $M_r<-20.0$ and $-21.0$,
  respectively, from the DR4 sample of \citet{baldry06}. The
  dash-and-dotted lines show the volume averages for $z<0.05$ ($=1$ by
  design) and $z<0.1$ ($=1.13$).}
\label{fig:zdist-density}
\end{figure}

While all the GSMFs are relatively similar, they all show a subtle
`dip' relative to a smooth slope around $10^{8.6}\Msun$ in
Fig.~\ref{fig:compare-gsmf-new}.  This feature is a product of using
$1/\Vmax$ with large-scale structure (LSS) variations
\citep*{EEP88}. Figure~\ref{fig:zdist-density} shows how the number
density of a volume-limited sample of galaxies ($M_r<-19$) varies with
redshift. There is a clear underdensity around $z\sim0.013$ that is
the cause of the GSMF dip. See also \citet{blanton05}, e.g.\ fig.~9 of
that paper, for the effect of this LSS on the luminosity function. A
simple way of removing this effect is to set $w_i$, for each galaxy,
equal to $1/n(z)$ where $n$ is the normalised number density as shown
in Fig.~\ref{fig:zdist-density} at the redshift of the galaxy. Before
recomputing the GSMF in this way, we discuss the selection limits
related to surface brightness and M-L ratio.  The mass used for
each galaxy in the rest of the paper is obtained from the mean of
the $\log \mass$ estimates.

\subsection{Selection limits}
\label{sec:selection-limits}

One of the key factors in considering the determination of the GSMF is
the SB limit, which is often not explicitly identified, of a redshift
survey \citep{CD02}. From the tests of \citet{blanton05}, as shown in
fig.~3 of that paper, the SDSS main galaxy sample has 70\% or greater
completeness in the range 18--$23{\rm\,mag\,arcsec}^{-2}$ for the
effective SB $\mu_{R50,r}$.

\begin{figure}
  \includegraphics[width=\singlecolsize\textwidth]{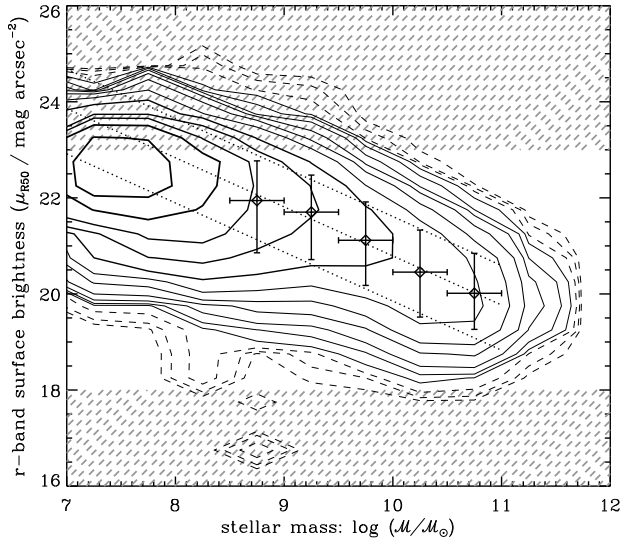}
\caption{Bivariate distribution for SB versus mass.  The contours
  represent the volume-corrected number densities from the sample:
  logarithmically spaced with four contours per factor of ten. 
  The lowest dashed contour corresponds to 
  $10^{-5}{\rm\,Mpc}^{-3}$ per $0.5\times0.5$ bin, while the lowest 
  solid contour corresponds to $5.6 \times 10^{-5}{\rm\,Mpc}^{-3}$. The
  grey dashed-line regions represent areas of low completeness (70\%
  or lower as estimated by \citealt{blanton05}). The diamonds with
  error bars represent the median and 1-sigma ranges over certain
  masses with a straight-line fit shown by the middle dotted line. The
  outer dotted lines represent $\pm1$ sigma.}
\label{fig:mass-sb}
\end{figure}

In order to identify at what point the GSMF becomes incomplete because
of the SB limit we computed the bivariate distribution in SB versus
stellar mass. Figure~\ref{fig:mass-sb} shows this distribution
represented by solid and dashed contours ($1/\Vmax$ and $1/n$ LSS
corrected). There is a relationship between peak SB and $\log \mass$,
which is approximately linear in the range $10^{8.5}$ to
$10^{11}\Msun$. At lower masses, the distribution is clearly affected
by the low SB incompleteness at
$\mu_{R50,r}>23{\rm\,mag\,arcsec}^{-2}$. Therefore, any GSMF values
for lower masses should be regarded as lower limits if there are no
corrections for SB completeness.

\begin{figure}
  \includegraphics[width=\singlecolsize\textwidth]{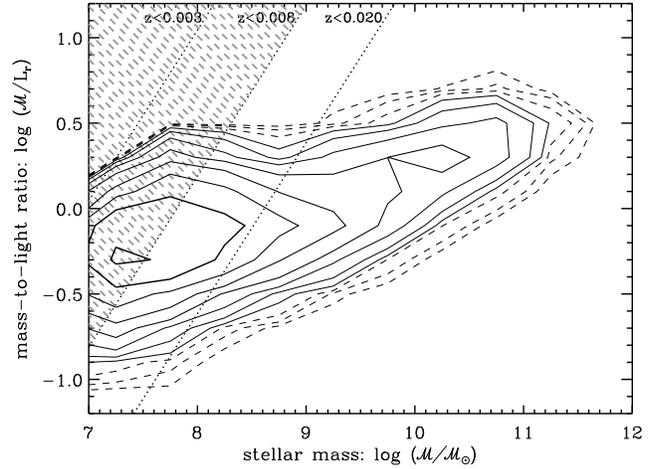}
\caption{Bivariate distribution for M-L ratio versus mass.  The
  contours represent the volume-corrected number densities:
  logarithmically spaced with four contours per factor of ten.  
  The lowest dashed contour corresponds to 
  $10^{-4}{\rm\,Mpc}^{-3}$ per $0.5\times0.2$ bin. The
  dotted lines represent the observable limits for an $r<17.8$
  magnitude limit and different redshift limits (ignoring
  k-corrections). The grey dashed-line region represents galaxies that
  can only be observed at $z<0.008$ where Hubble-flow corrections are
  significant ($cz<2400{\rm\,km\,s}^{-1}$).}
\label{fig:mass-ml}
\end{figure}

The other important consideration is fact that the $r$-band selection is not
identical to the mass selection required for the GSMF. This is nominally
corrected for by $1/\Vmax$ but it should be noted that galaxies with high M-L
ratio at a given mass are viewed over significantly smaller volumes than those
with low M-L ratio.  Figure~\ref{fig:mass-ml} shows the bivariate distribution
in M-L ratio versus mass. The limits at various redshifts for the SDSS main
galaxy sample are also identified. For example, galaxies with $\mass <
10^{8}\Msun$ and $\log(\mass/L_r) > 0.1$ are only in the sample at
$z<0.008$. At these low redshifts, the stellar mass and $\Vmax$ depend
significantly on the Hubble-flow corrections. However, it does appear that the
SB limit affects the completeness of GSMF values at higher masses than the M-L
limits. At $\mass < 10^{8.5}\Msun$ the SB limit becomes significant, while at
$\mass < 10^{8}\Msun$, the GSMF is affected both by the constrained volume for
high M-L ratio galaxies and more severely by the SB incompleteness.

\subsection{Corrected GSMF with lower limits at the faint end}
\label{sec:final-gsmf}

\begin{figure}
  \includegraphics[width=\singlecolsize\textwidth]{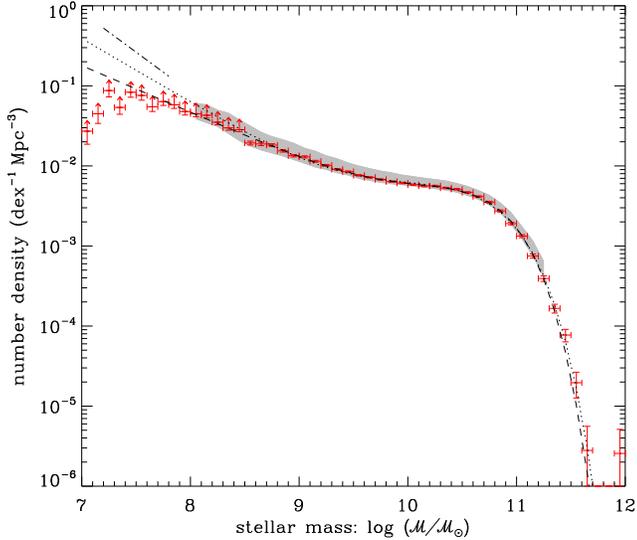}
\caption{GSMF extending to down to $10^{7}\Msun$ determined from the
  NYU-VAGC. The points represent the non-parametric GSMF with Poisson
  error bars; at $\mass < 10^{8.5}\Msun$ the data are shown as lower
  limits because of the SB incompleteness
  (Fig.~\ref{fig:mass-sb}). The dashed line represents a
  double-Schechter function extrapolated from a fit to the $\mass >
  10^{8}\Msun$ data points.  The dotted line shows the same type of
  function with a faint-end slope of $\alpha_2 = -1.8$ (fitted to
  $\mass > 10^{8.5}\Msun$ data). The dash-and-dotted line represents a
  power-law slope of $-2.0$. The shaded region shows the range in the
  GSMF from varying the stellar mass used and changing the redshift
  range.}
\label{fig:gsmf-special}
\end{figure}

Figure~\ref{fig:gsmf-special} shows the results of the GSMF
determination. The binned GSMF is represented by points with Poisson
error bars, with lower limits represented by arrows. The GSMF has been
corrected for volume ($1/\Vmax$) and LSS ($1/n$) effects.\footnote{We
compared the GSMF computed using $1/n(z)$ correction for LSS
variations with the stepwise maximum-likelihood method \citep{EEP88}.
There was good agreement between the two methods after matching
normalisations.  The former method is simpler when computing bivariate
distributions.}  The masses used were the average of the four
estimated stellar masses. The shaded region represents the full range
in the GSMF obtained by varying the stellar mass used
(Fig.~\ref{fig:compare-gsmf-new}) and multiplying the upper number
densities by 1.13 to account for the expected renormalisation at
$z<0.1$ (Fig.~\ref{fig:zdist-density}).  After renormalisation, the
range in $f_s$ is 0.038--0.048. In addition if we multiply by 1.2 to
account for missing light (optically-thick regions and corrections to
total magnitudes, \S~\ref{sec:compare-gsmf}), we obtain 0.046--0.058.

The binned data for the GSMF can clearly not be represented
by a single \citet{Schechter76} function. The data were fitted with
a double Schechter function given by
\begin{equation}
  \phi_\mass \, \D \mass = e^{-\mass/\mass^{*}} \left[ \phi^{*}_1 \left(
  \frac{\mass}{\mass^{*}} \right)^{\alpha_1} + \phi^{*}_2 \left(
  \frac{\mass}{\mass^{*}} \right)^{\alpha_2} \right] \, 
  \frac{ \D \mass }{ \mass^{*} }
\label{eqn:double-schechter}
\end{equation}
where $\phi_\mass \, \D \mass$ is the number density of galaxies with
mass between $\mass$ and $\mass + \D \mass$; with $\alpha_2 <
\alpha_1$ so that the second term dominates at the faintest
magnitudes. Fitting to $\mass>10^{8}\Msun$, the best-fit parameters are
\begin{equation}
\begin{array}{ll}
\log (\mass^{*}/\Msun) = 10.648 \\
\phi^{*}_1 / 10^{-3}{\rm\,Mpc}^{-3} = 4.26 & \alpha_1 = -0.46 \\ 
\phi^{*}_2 / 10^{-3}{\rm\,Mpc}^{-3} = 0.58 & \alpha_2 = -1.58
\end{array}
\label{eqn:d-schechter-paras}
\end{equation}
with formal errors of 0.013, 0.09, 0.05, 0.07, 0.02. The dashed line
in Fig.~\ref{fig:gsmf-special} represents this fit extrapolated down
to $10^{7}\Msun$.
%%  Parameters for Double Schechter fit (13 DEC 2007)
%%    10.648    0.00426      -0.46   0.00058      -1.58   
%%     0.013    0.00009       0.05   0.00007       0.02

Even though the Poisson errors are small, for illustrative purposes
and because systematic errors are clearly significant, we fitted a
function with $\alpha_2=-1.8$ fixed. This is represented by the dotted
line in Fig.~\ref{fig:gsmf-special} and to the eye provides an equally
good fit to the data at $\mass > 10^{8.6} \Msun$. Given the SB
incompleteness (Fig.~\ref{fig:mass-sb}), a steep faint-end slope such
as this cannot be ruled out. 

\subsection{Comparison with cluster environments}
\label{sec:discuss-gsmf}

The field GSMF shows a clear signal of a change in slope at masses
lower than the characteristic mass;\footnote{It makes minimal
difference to the shape of the cosmic volume average GSMF, and no
difference to the discussion in this paper, if the highest density
regions (15\% of the population, i.e.\ clusters and compact groups)
are excluded from the calculation.  This justifies the use of the term
`field' to describe this GSMF.} as was already evident in the
luminosity function of the redder SDSS bands \citep{blanton05}. Thus
there is a significant difference between a faint-end slope determined
from a Schechter fit around the characteristic mass (luminosity) and
the faint-end slope at lower masses (luminosities).

Recently, \citet{popesso06} and \citet{jenkins07} have confirmed earlier
reports of upturns in the faint end of cluster luminosity functions.  These
were based on the RASS-SDSS galaxy cluster survey and 3.6-$\mu{\rm m}$ imaging
of the Coma Cluster using the Infrared Array Camera on the Spitzer Space
Telescope, respectively.

\begin{figure}
  \includegraphics[width=\singlecolsize\textwidth]{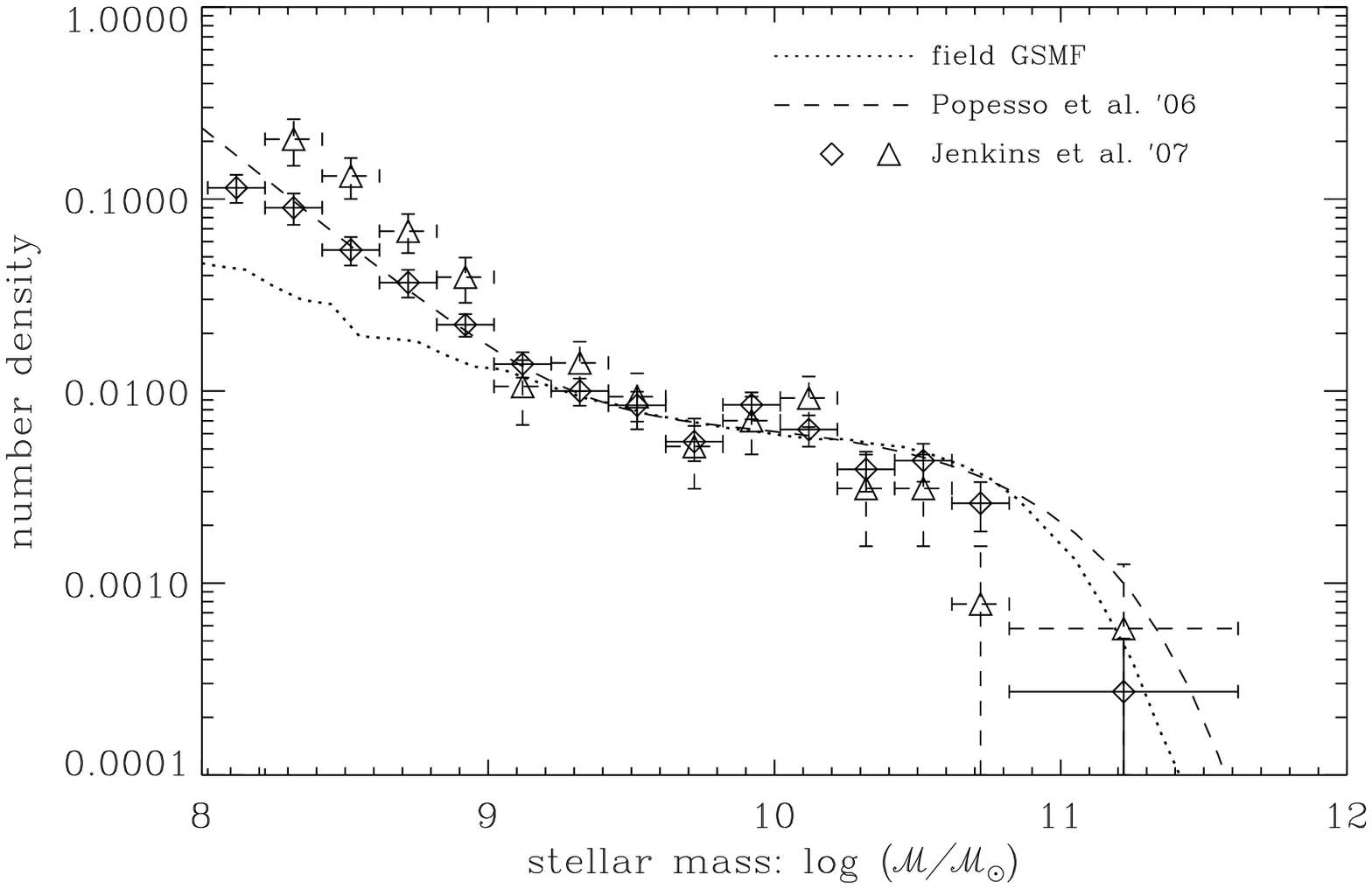}
\caption{Comparison of inferred GSMFs from the Coma Cluster
  3.6-$\mu{\rm m}$ survey of \citeauthor{jenkins07}\ and the SDSS cluster
  survey of \citeauthor{popesso06}\ with the field GSMF (dotted line). The
  diamonds and triangles represent the Coma core and Coma off-centre
  columns from table~2 of \citeauthor{jenkins07}, using the $g-r<1.3$
  restriction for the faint end.  The dashed line represents the
  \citeauthor{popesso06} fit derived from the $z$-band luminosity
  function.  The data were normalised to match the field GSMF around
  $\log \mass_s = 10.0 \pm 0.5$.}
\label{fig:compare-gsmf-cluster}
\end{figure}

Stellar M-L ratio variations between cluster galaxies are typically
less severe than between field galaxies. Using a simple conversion
between absolute magnitude and stellar mass given by
$
  \log \mass_s = (M_{\rm solar} - M)/2.5 + \log(\mass_s/ L) \, ,
$
we converted the cluster luminosity functions to GSMFs: with $M_{z,{\rm
solar}} = 4.4$ (AB mags), $\log(\mass_s/ L_z) = 0.2$ (solar units),
$M_{3.6,{\rm solar}} = 3.3$ (Vega mags) and $\log(\mass_s/ L_{3.6}) = -0.5$
(solar units).  The conversion factors were estimated using \pegase\ and the
filter curves. Figure~\ref{fig:compare-gsmf-cluster} shows the resulting
cluster GSMFs.  The faint-end upturn is evident at $\mass_s < 10^{9} \Msun$
and is significantly steeper than the field GSMF.  Thus it appears that the
slope of the GSMF around $10^{8}$ to $10^{9}\Msun$ depends significantly on
the environment.  Note however that these cluster results rely on estimated
subtraction of background galaxy counts.

As well as the difference in the slope of the GSMF in the range
$10^{8}$ to $10^{9}\Msun$, there is the more established difference
between the morphologies of these low-mass galaxies in different
environments. In clusters, these are predominantly dwarf ellipticals
(dE); whereas in the field, these are predominantly late-type spirals
(Sd) and irregulars (Sm,Im) (e.g.\ by estimating stellar masses for
galaxies in the Nearby Field Galaxy Survey of \citealt{jansen00}).
Field low-mass galaxies are generally forming stars and have
substantial reservoirs of gas \citep{swaters02}, and therefore their
baryonic masses can be several times their stellar masses.

\section{The stellar mass-metallicity relation and the baryonic mass function}
\label{sec:mzr}

In order to convert our stellar mass function (MF) to the more fundamental
baryonic MF, we develop a method for deriving the stellar mass fraction (i.e.,
conversion factor of baryonic mass to stars) in terms of the stellar
mass. This can be achieved by using the well established relation between
stellar mass and metallicity coupled with a metallicity to stellar mass
fraction relation, which can be determined from a simple model. This method is
laid out below.  We assume the following: (i) gas in a galaxy is well
mixed, in particular, we do not include metal-enriched outflows in the model;
(ii) the galaxy gas-phase metallicity measured from an SDSS spectrum
represents an effective average over the whole galaxy and can be related to
the global gas fraction, i.e., there is no consideration of metallicity and
related gradients in galaxies. Despite these simplifications, the derived
average stellar mass fractions are shown to be consistent with direct measures
of gas masses.

\subsection{Relating metallicity to stellar mass fraction}

The importance of stellar mass is demonstrated by the tight relationship
between gas-phase metallicity and stellar mass \citep{tremonti04}, and the
relationship between metallicity and the fraction of baryonic mass in stars.
The latter is illustrated using a three-component model with stellar mass,
retained gas and expelled gas, which reduces to the closed-box model in the
case of zero expelled gas.  Regardless of the problems with this simple model,
there clearly must be a fundamental relationship between the fraction of
baryons locked up in stars, gas and the progress of chemical evolution in a
galaxy.

From eq.~5 of \citet{Edmunds90} with a simple outflow that is
proportional to the SF rate and no infall, we can set
\begin{eqnarray}
 \D (gZ) & = & (a p  - a Z  - o Z) \, \D s \\
 \D g    & = & - (a + o) \, \D s 
\end{eqnarray}
where: $g$ is the mass of gas, $Z$ is the metallicity of the gas, $p$
is the yield defined as the mass of metals produced per mass of
long-lived stars formed, $s$ is the integrated mass of stars formed,
$a$ is the fraction of mass in stars that is not instantly recycled,
i.e.\ $a s$ is the stellar mass of the galaxy, and $o$ is the mass of
gas expelled per mass of stars formed. The integral is then given by
\begin{eqnarray}
  \D Z & = & - \left( \frac{a p}{a + o} \right)  \frac{1}{g} \, \D g \\
  Z    & = &   \left( \frac{a p}{a + o} \right)  
               \ln \left( \frac{g_{\rm i}}{g} \right) 
\label{eqn:Z-outflow-model}
\end{eqnarray}
where $g_{\rm i}$ is the initial gas mass, i.e., the total mass of
stars and gas (the baryonic mass), and $g = g_{\rm i} - a s - o s\:$.
Setting $o=0$, Eq.~\ref{eqn:Z-outflow-model} reduces to the standard
closed-box solution given by
\begin{eqnarray}
     Z & = & p \ln \left( \frac{g_{\rm i}}{g} \right) 
\: \equiv \: - p \ln \left( 1 - \frac{a s}{g_{\rm i}} \right) \: .
\label{eqn:closed-box}
\end{eqnarray}

\begin{figure}
  \includegraphics[width=\singlecolsize\textwidth]{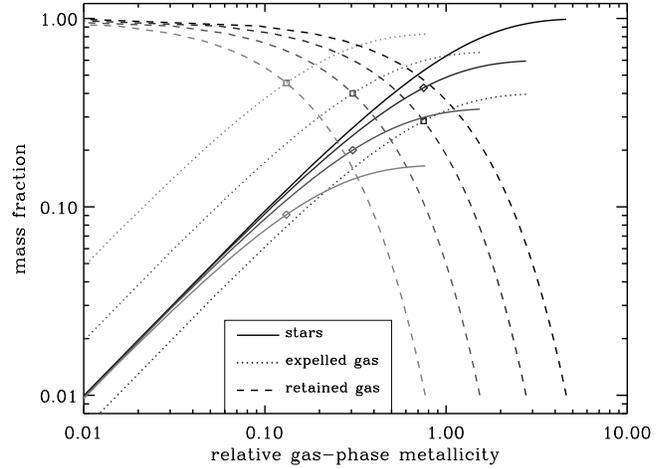}
\caption{The mass fraction of stars (solid lines), retained
  inter-stellar gas (dashed lines) and expelled gas (dotted lines)
  versus the gas-phase metallicity for the three-component model
  (Eq.~\ref{eqn:Z-outflow-model}). The darkest lines represent a
  closed-box model with no outflow while the lighter lines represent
  increasing outflow factors (0, 0.4, 1.2, 3.0). The diamonds and
  squares represent points at which the mass in expelled gas is equal
  to the mass in retained gas.}
\label{fig:metal-mass-model}
\end{figure}

Figure~\ref{fig:metal-mass-model} shows how the fraction of mass in
stars ($a s/g_i$), retained inter-stellar gas ($g/g_i$) and
expelled gas ($os/g_i$) depends on the metallicity.  The curves were
obtained from Eq.~\ref{eqn:Z-outflow-model} with $a=0.6$ and
$o=0.0,0.4,1.2,3.0$. The curves are shown for $g/g_i \ge 0.01$ i.e.\ a
minimum of 1\% of the mass remaining as inter-stellar gas.

The simple model demonstrates that the gas-phase metallicity can be a
relatively accurate predictor of a galaxy's SF efficiency ($a s/g_i$ or
baryonic-to-stellar mass conversion factor) regardless of a wide range of
outflow scenarios.  If the retained gas is greater in mass than the expelled
gas, the SF efficiency will be within 30\% of the closed-box estimate derived
from the metallicity regardless of the outflow factor.  If there is
significant expelled gas, using Eq.~\ref{eqn:closed-box} to relate metallicity
to the stellar mass fraction will result in an underestimate of $a s / (a s +
g)$ and an overestimate of $a s / g_i$. If there are continuing infalls and
outflows (e.g.\ \citealt{Dalcanton07,Erb08}), the case is less clear because of
the dependence on the metallicity of the infalling gas.

\begin{figure}
  \includegraphics[width=\smallcolsize\textwidth]{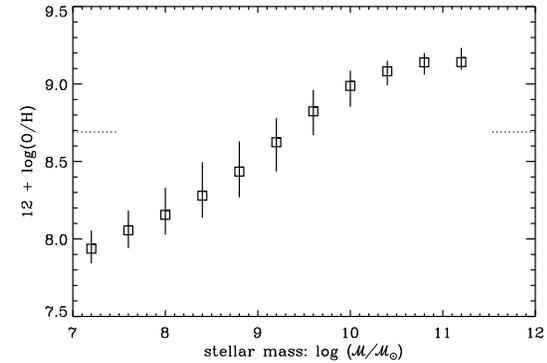}
\caption{The stellar mass-metallicity relation derived from the
  emission-line analysis by \citet{tremonti04}. The squares represent
  the median while the lines represent the 16th and 84th percentiles
  in logarithmic mass bins. The dotted lines show the
  solar abundance value determined by \citet{allende01}.}
\label{fig:metal-mass-data}
\end{figure}

Figure~\ref{fig:metal-mass-data} shows stellar mass-metallicity relation using
gas-phase oxygen abundances estimated by \citet{tremonti04}. This does not
vary significantly with environment \citep*{MBB07} suggesting that there is a
fundamental relationship between the present-day stellar mass of a galaxy and
its SF efficiency. \citet{brooks07} concluded from their modelling that ``low
star formation efficiencies ...  are primarily responsible for the lower
metallicities of low-mass galaxies and the overall $\mass$-$Z$ trend''.

Note that the estimated metallicity can depend on the emission lines
considered and the calibration
\citep{KK04metallicity,savaglio05,KE08}.  However our consideration
here is only that the measured $\mass$-$Z$ relation implies an
increase of average gas-phase metallicity with mass, and that the
relation between metallicity and stellar mass is fairly tight.

\subsection{Relating stellar mass to baryonic mass}

\begin{figure}
  \includegraphics[width=\singlecolsize\textwidth]{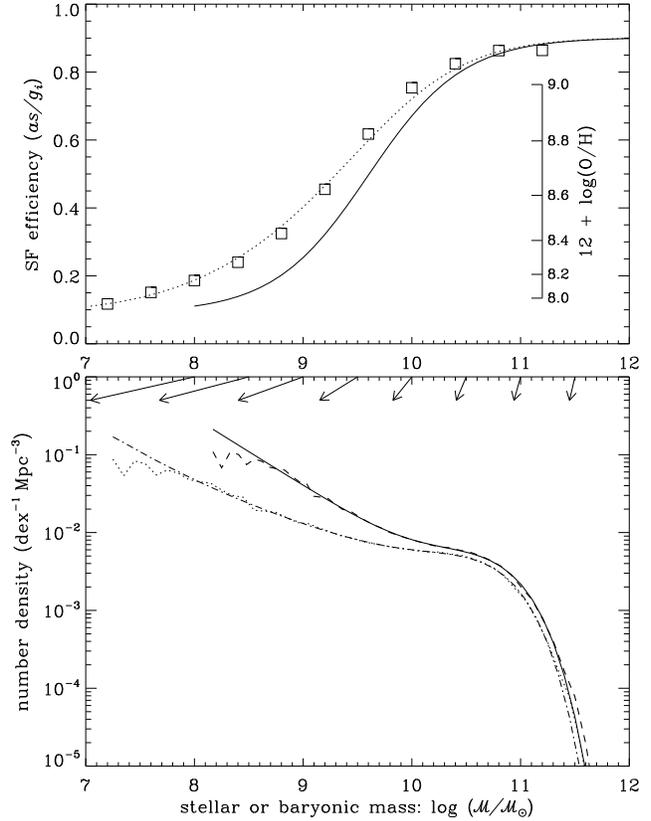}
\caption{SF efficiency versus mass and the implied baryonic mass
  function. {\em Upper panel}: The model SF efficiency is defined as
  the mass in stars formed divided by the mass available to form stars
  over the life of a galaxy. This is plotted versus stellar mass
  (dotted line) and baryonic mass (solid line).  The squares represent
  the O/H values of Fig.~\ref{fig:metal-mass-data} converted using a
  monotonic relationship to efficiency determined from a closed-box
  model and plotted against stellar mass.  {\em Lower panel}: The
  dotted line represents the GSMF from
  Fig.~\ref{fig:gsmf-special}. The dashed line represents the implied
  baryonic mass function given the relationship between SF efficiency
  and stellar mass. The solid line represents a double Schechter fit
  to these data at $\mass_b > 10^{9}\Msun$ with faint-end slope
  $\alpha_2\simeq-1.9$. The dash-and-dotted line is this fit converted
  back to stellar mass. The arrows illustrate the relationship between
  baryonic and stellar mass in this simple model.}
\label{fig:implied-bf}
\end{figure}

The variation in SF efficiency implied by the $\mass$-$Z$ relation
allows one to estimate a baryonic MF.  We assumed (i) the
median measured oxygen abundances at a given stellar mass are
representative of the average gas-phase metallicity in the
inter-stellar medium, (ii) the yield is independent of galaxy mass and
time, (iii) the closed-box model can be used to relate SF efficiency
to oxygen abundance, and (iv) the SF efficiencies of the most massive
galaxies are about 90\%. Next, we defined a parametrisation to relate
the total baryonic mass of a galaxy ($\mass_b$) to the SF efficiency: 
\begin{equation}
\frac{a s}{g_i}  \equiv  \frac{\mass_s}{\mass_b} = 
  \frac{\mass_b}{\mass_b + \mass_0} (e_u - e_l) + e_l
\label{eqn:sf-para}
\end{equation}
where $\mass_0$ is a characteristic mass, and $e_u$ and $e_l$ are 
upper and lower limits of the SF efficiency. Combining with
Eq.~\ref{eqn:closed-box}, this parametrisation was fitted to the
$\mass_s$-$Z$ data (using O/H, Fig.~\ref{fig:metal-mass-data}, for
$Z$).  Figure~\ref{fig:implied-bf} shows the SF efficiency versus
$\log \mass_s$ and $\log \mass_b$ from the best-fit parametrisation
($\log \mass_o=9.6$, $e_u=0.90$, $e_l=0.09$, $12 + \log p = 8.84$
where $p$ is the oxygen yield).

The implied relationship between stellar and baryonic mass was then
used to compute the baryonic MF (from the GSMF or from the
stellar masses of all the galaxies). This is shown by the dashed line
in the lower panel of Fig.~\ref{fig:implied-bf}.  The solid line is a
fit to the MF at $\mass_b > 10^{8.5} \Msun$.  The faint-end
slope in this case is given by a steep $\alpha_2\simeq-1.9$  (parameters
as per Eq.~\ref{eqn:d-schechter-paras} are 10.675, 4.90, $-0.42$, 0.61,
$-1.87$.) 
%%  Parameters for Double Schechter fit (13 DEC 2007)
%%    10.675    0.00490      -0.42   0.00061      -1.87 
%%     0.014    0.00010       0.05   0.00007       0.02 

The steep part of the baryonic MF is caused by the initially fairly
steep GSMF at $\mass_s < 10^{9}\Msun$ coupled with an increase in
gas-to-stellar (G-S) mass ratio from $\sim1.5$ at $10^{9}\Msun$ to $\sim4$
at $10^{8}\Msun$ to $\sim8$ at $10^{7}\Msun$
(Fig.~\ref{fig:implied-bf}: dotted line in upper panel).  
Though the G-S mass ratios for low-mass galaxies
are large they are in fact similar to those obtained directly from
\HI\ surveys.  For example, the above values are similar to the
ratios estimated by \citet{kannappan04} for the low-mass end of the
blue sequence.

In order to test the average G-S mass ratios implied by our
parametrisation, we used stellar and atomic gas masses derived from
the Westerbork \HI\ Survey \citep{SB02,noordermeer05} and the
literature compilation of \citet{Garnett02}. The stellar masses were
estimated using the simple relations of \citet{bell03} applied to the
$B$ and $R$ photometry of \citeauthor{SB02} and the $B$ and $V$
photometry of \citeauthor{Garnett02}, and using $\log(\mass_s/ L_B) =
0.3$ for the early-type spirals of \citeauthor{noordermeer05} We also
matched the \HI\ Parkes All-Sky Survey (HIPASS) catalogue
\citep{meyer04,wong06} to the NYU-VAGC catalogue.  This resulted in
170 galaxies with one-and-only-one match within $0.25\degr$ and
$\Delta z c < 250{\rm\,km\,s}^{-1}$.  

\begin{figure}
  \includegraphics[width=\singlecolsize\textwidth]{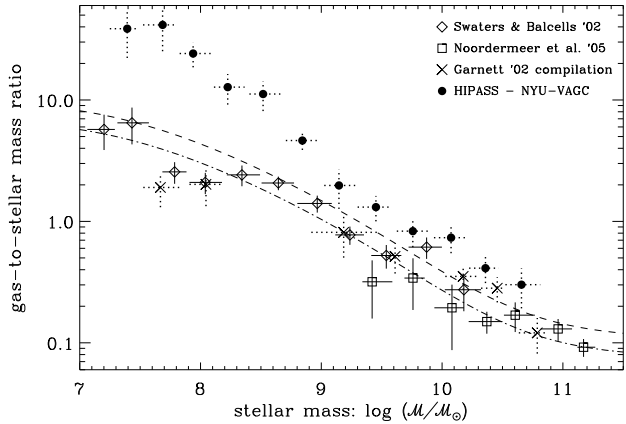}
  \caption{Average atomic G-S mass ratios. The symbols represent G-S mass
  ratios from \HIcaption\ measurements with a 1.33 correction factor to
  account for Helium. The data from 4--25 galaxies were averaged for each
  point.  The vertical error bars represent the standard errors; while the
  horizontal error bars represent the ranges of each sample. See text for
  further details. The dashed line represents the parametrisation of
  Eq.~\ref{eqn:sf-para} assuming the closed-box model and that all the gas is
  atomic gas.  The dash-and-dotted line represents the same parametrisation
  but assuming 70\% of the gas is atomic gas. Note that gas-poor ellipticals
  are not considered in these samples, however, they are only a significant
  field population at $\ga 10^{11}\Msun$ and so their gas properties can be
  ignored in terms of computing the baryonic MF.}
\label{fig:gs-ratio}
\end{figure}

Figure~\ref{fig:gs-ratio} shows average G-S mass ratios in bins of
stellar mass for these surveys; also shown is the relation derived
from our parametrisation.  The HIPASS mass ratios lie above this line,
not surprisingly, because the \HI\ selection misses galaxies with low
G-S mass ratios. Thus these derived mass ratios should be regarded as
upper limits.  The G-S mass ratios derived from the optically-selected
Westerbork survey are in good agreement, which lends support to our
parametrisation of the SF efficiency. The Garnett compilation points
to a flatter relation but this was using a significantly smaller,
heterogeneously-selected sample.

Various authors have found that the effective yield is lower in low mass
galaxies \citep{Garnett02,PVC04,tremonti04} with an interpretation being that
there is significant metal loss by metal-enriched outflows
\citep{Dalcanton07}.  The effective yield is defined as
\begin{equation}
p_{\rm eff} = Z / \ln \left( \frac{as+g}{g} \right)
\end{equation}
such that $p_{\rm eff} = p$ for the closed-box model, and $p_{\rm eff}<p$ with
outflows.  Our analysis uses SDSS-aperture metallicities and a closed-box
model to predict average gas masses. This gives approximately correct global
G-S mass ratios at least in comparison with the Westerbork samples
(Fig.~\ref{fig:gs-ratio}), which is consistent with the variation in SF
efficiency with mass being the primary cause of the $\mass_s$-$Z$ relation
(see also \citealt{brooks07,ellison08,mouhcine08,TKG08}). 
This does not mean that there
are no significant outflows, only that they may be a secondary cause of the
$\mass_s$-$Z$ relation.

The precise shape and steepness of the baryonic MF does depend on the detail
of the G-S mass ratios and the dispersion in the relationship with stellar
mass. However, our main aim is to illustrate the relationship between galaxy
baryonic mass function and GSMF, and so we use the simple stellar-to-baryonic
mass relation derived here. We also note that while the change in slope
converting from the GSMF to the baryonic MF could be exaggerated, the GSMF
slope may be underestimated and so the faint-end slope of the baryonic MF
could be $\sim-1.9$ even accounting for a more complicated relation.

\subsection{Comparison between galaxy and halo baryonic mass functions}
\label{sec:discuss-bf}

The implied baryonic mass function (Fig.~\ref{fig:implied-bf}) can be
considered to be the sum of stars and any gas involved with the cycle of SF,
in and around each galaxy, as long as the relationship between
$\mass_s/\mass_b$ and $Z$ is not strongly affected by outflows. Other
estimates of the baryonic MF have been made by \citet{bell03gbmf}, 
\citet{RT05} and \citet{shankar06}. 
These include stars and atomic gas in galaxies, 
and molecular gas in the case of the first two estimates.

\begin{figure}
  \includegraphics[width=\singlecolsize\textwidth]{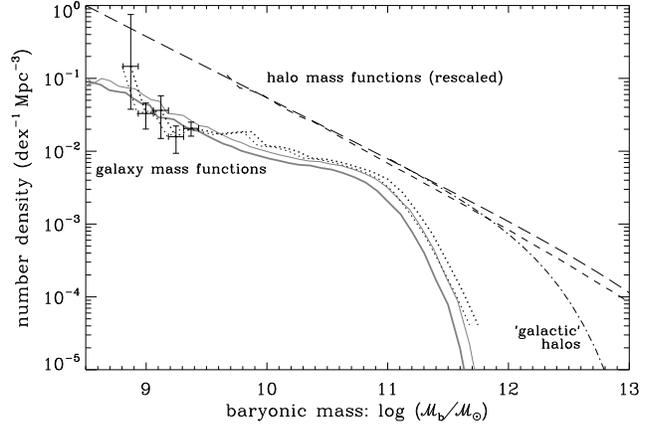}
  \caption{Comparison between galaxy and halo baryonic mass functions.
    The thick dotted line represents the MF determined by
    \citet{bell03gbmf}; with the thin dotted line representing masses
    reduced by 15\%. Error bars are shown for the low-mass end.  The
    thick solid line represents the MF from Fig.~\ref{fig:implied-bf};
    with the thin line representing masses increased by 20\% and
    number densities by 13\% (cf.\ \S~\ref{sec:final-gsmf}). Halo MFs
    with masses multiplied by $\Omega_b/\Omega_m$ are shown by
    the dashed lines (Millennium Simulation; \citealt{ST99}
    using longer dashes), and by the dash-and-dotted line for the
    `galactic halo MF' \citep{shankar06}.}
\label{fig:bmf-compare}
\end{figure}

Figure~\ref{fig:bmf-compare} shows a comparison between the implied
baryonic mass function and the diet Salpeter version from
\citet{bell03gbmf}. There is a significant offset between them, which
can be reconciled by (i) allowing for the number density and missing
light corrections of \S~\ref{sec:final-gsmf} and (ii) slightly
lowering the masses of \citeauthor{bell03gbmf}\ because the stellar
M-L ratios appear on the high side (\S~\ref{sec:compare-gsmf}).  This
is partly because of using the diet Salpeter IMF compared to those of
Kroupa or Chabrier (Appendix). After plausible corrections, the galaxy
baryonic MFs are in good agreement.  There is no clear evidence for a
steep slope at $\mass \la 10^{9.5}\Msun$ in the
\citeauthor{bell03gbmf}\ function, but given the large error bars, it
is consistent with the steep slope we find.  Neither
\citet{RT05} or \citet{shankar06} noted such a steep slope in the
baryonic MF. \citeauthor{shankar06}'s G-S mass ratios were based on a
calibration of \HI\ and stellar masses as a function $B$-band luminosity
(from \citealt{SP99}). Their
average G-S mass ratio is $\sim1$ at $\mass_s=10^8\Msun$, which is
significantly lower than our estimate.

For comparison, halo MFs are shown in
Fig.~\ref{fig:bmf-compare} assuming a constant fraction of mass in
baryons in each halo ($\Omega_b/\Omega_m$). The long-dashed line
was derived from the simulations of \citet{ST99} while the dashed line
was derived from the Millennium Simulation
\citep{springel05nat}.\footnote{The Millennium Simulation is available
at \\ http://www.mpa-garching.mpg.de/millennium/} The SF efficiency
implied by the $\mass_s$-$Z$ relation shows that the galaxy baryonic
MF is approximately as steep, at $\mass_b < 10^{9.5}
\Msun$, as the halo MF. 
The `galactic halo MF' given by eq.~9 of \citet{shankar06}
is also shown.  This galactic halo MF, which includes halos and
sub-halos hosting a galaxy but not group and cluster halos, is very
similar in shape to the standard halo MF at $\mass_b \la
10^{11}\Msun$.  This is because the slopes of sub-halo MFs do not
depend strongly on the mass of the main halo and have values similar
to that of the halo MF \citep{delucia04shmf,reed05}.

\begin{figure}
  \includegraphics[width=\singlecolsize\textwidth]{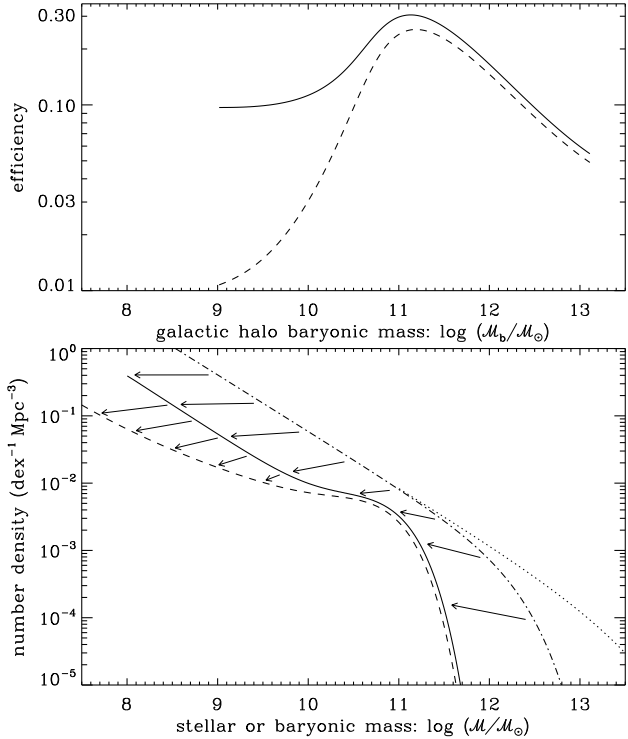}
\caption{Idealised reconstruction of galaxy formation efficiency.  The
  top panel shows the fraction of baryons falling onto a galaxy (solid
  line) and the fraction of baryons forming stars (dashed line).  The
  lower panel shows the result of using the efficiency curves to go
  from the galactic halo baryonic MF (dash-and-dotted line)
  to the galaxy baryonic MF (solid line) and to the GSMF (dashed
  line).  A standard halo baryonic MF that does not distinguish
  between galactic, group and cluster halos is shown by the dotted
  line.  The arrows illustrate the relationships between the mass
  functions.  They are not flat because the number density per
  logarithmic bin changes depending on the derivative of efficiency
  with respect to mass.}
\label{fig:halo-eff}
\end{figure}

To illustrate the implications of the form of the mass functions on
galaxy formation efficiency, we computed the efficiency as a function
of halo baryonic mass required to reproduce the galaxy MFs.
This assumes a one-to-one and monotonic relationship between
halo mass and galaxy mass. In detail this is described by
\citet{shankar06}, and earlier using galaxy luminosity by \citet{MH02}
and \citet{VO04}.  For the halo MF, we use the galactic halo MF
with mass multiplied by $\Omega_b/\Omega_m$.
Figure~\ref{fig:halo-eff} shows the efficiency versus halo mass and
the reconstructed galaxy MFs. The efficiency represented by the solid
line can be regarded as the fraction of baryons that fall onto a
galaxy and are available for forming stars; while the dashed line
represents the fraction of baryons that have formed stars.  These
reproduce the galaxy baryonic MF and GSMF, respectively.

The figure demonstrates the implication that there is a
levelling off of the baryonic-infall efficiency at low masses, while
SF efficiency continues to fall, and shows a peak galaxy formation
efficiency at $10^{11.1}\Msun$ ($10^{11.9}\Msun$ including CDM).
\citet{shankar06} also looked at SF efficiency versus halo mass
and demonstrated a similar result (their fig.~5) albeit with a flatter
falloff at higher masses and a steeper falloff at lower masses related
to the different GSMF utilised by them.
The reduced efficiency above and below this
scale are inherent in recent semi-analytical models of galaxy
formation including SF and active galactic nuclei feedback
\citep{bower06,cattaneo06,croton06}. The scale is related to a minimum
in feedback effects (cf.\ fig.~8 of \citealt{DB06}).

\section{Summary and conclusions}
\label{sec:conclusions}

The field low-redshift GSMF has been generally measured down to $\sim
10^{8.5} \Msun$ (Fig.~\ref{fig:compare-gsmf}). This accounts for the
majority of stellar mass in the universe.  The SMD of the universe is
in the range 5--7 per cent of the baryon density (conservatively 4--8
per cent, assuming $\Omega_b=0.045$).  The sources of uncertainty are
corrections for dust attenuation, underlying stellar M-L ratios,
missing low-SB light, and normalisation of the GSMF.

We determined the field GSMF using the SDSS NYU-VAGC low-redshift
galaxy sample.  At masses below $\sim10^{9}\Msun$, the shape of the
GSMF computed using a standard $1/\Vmax$ approach was affected by LSS
variations. This was corrected for using number density as a function
of redshift of a volume-limited sample of galaxies
(Fig.~\ref{fig:zdist-density}). Analysis of the bivariate distribution
of stellar mass versus SB indicated that the number densities should
be regarded as lower limits at masses below $\sim10^{8.5}\Msun$
(Fig.~\ref{fig:mass-sb}).  Despite this incompleteness, there is clear
evidence for an upturn in the GSMF (Fig.~\ref{fig:gsmf-special}) with
a faint-end slope $\alpha_2 \simeq -1.6$
(Eq.~\ref{eqn:d-schechter-paras}). This represents the power-law slope
over the range $\la 10^8$ to $\sim 10^9\Msun$.  Steeper slopes are also
plausible. Slopes of $\sim -2$ have been measured in clusters over the
same mass range (Fig.~\ref{fig:compare-gsmf-cluster}).  The processes
shaping the GSMF in clusters will include stripping of gas and stars,
while feedback processes may be more dominant in shaping the field GSMF.

At masses below $10^{10}\Msun$, blue-sequence (late-type) galaxies are
the dominant field population (e.g.\ fig.~10 of \citealt{baldry04}).
The gas-phase metallicity of these galaxies can be related to a
SF efficiency that is defined as the total mass of stars
formed divided by the available baryonic mass for forming stars.  The
closed-box formula can remain a good estimate of the SF efficiency
even with moderate outflows (Fig.~\ref{fig:metal-mass-model}),
excluding the case of metal-enriched outflows for 
gas-rich systems \citep{Dalcanton07}.

Using a simple relationship between stellar mass and baryonic mass, based on
the $\mass_s$-$Z$ relation and the closed-box formula for $Z$ that
neglects outflows, we converted the field GSMF to an implied baryonic mass
function (Fig.~\ref{fig:implied-bf}). The resulting faint-end slope $\alpha_2
\sim -1.9$ is similar to the halo MF.  We note that this is only suggestive as
it depends on the form of the conversion between $\mass_s$ and $\mass_b$ and
the dispersion in this relationship (Fig.~\ref{fig:gs-ratio}). The shape of
the galaxy baryonic MF is consistent with the non-parametric MF of
\citet{bell03gbmf} (Fig.~\ref{fig:bmf-compare}).

Taking the shape of the implied baryonic mass function at face value,
we compared this with a simulated halo baryonic MF. Using a one-to-one
relationship between halos and galaxies, these can be used to
determine the galaxy formation efficiency (the fraction of baryons
falling onto a galaxy) as a function of halo mass
(Fig.~\ref{fig:halo-eff}). This illustrates how varying efficiency
with mass can be used to obtain galaxy mass functions from the halo
MF, or from a similarly-shaped sub-halo MF. The peak in the formation
efficiency curve may correspond to a minimum in feedback efficiency.

There is no evidence yet of any cutoff in mass, below which baryons do not
collapse into galaxies (cf.\ \citealt{DW03}). Rather we find the
baryonic-infall efficiency levels off to $\sim10$\% rather than continuing to
plummet with mass. It is possible that a cutoff mass scale, imprinted in the
shape of the field galaxy baryonic MF, could be found by future deeper
surveys. To robustly identify this scale requires: (i) wide-field deep imaging
with reliable identification of galaxies down to at least
$\mu_{r}\sim25{\rm\,mag\,arcsec}^2$; (ii) spectroscopy of large samples down
to $r\sim20{\rm\,mag}$, or a similar effective depth in near-IR selection, in
order to obtain accurate distances ($z\ga0.01$) and metallicities for low-mass
galaxies; (iii) wide-field \HI\ surveys in order to estimate gas masses more
directly.  The prospect of such a measurement within the next decade is good
with the advent of new wide-field optical instruments and survey-efficient
radio telescopes.

\section*{Acknowledgements}

We are grateful to Benjamin Panter for providing his stellar mass
catalogue, Phil James for comments on the paper, Maurizio Salaris for
\basti\ mass-to-light ratios, Rachel Somerville for halo mass
function data, and the anonymous referee for suggesting useful clarifications. 
I.K.B.\ acknowledges funding from the Science and
Technology Facilities Council. K.G.\ acknowledges funding from the
Australian Research Council.  We thank Michael Blanton and the authors
of the NYU-VAGC for publicly releasing their catalogue derived from
the SDSS.

Funding for the Sloan Digital Sky Survey has been provided by the Alfred
P. Sloan Foundation, the Participating Institutions, the National Aeronautics
and Space Administration, the National Science Foundation, the U.S. Department
of Energy, the Japanese Monbukagakusho, and the Max Planck Society. The SDSS
Web site is http://www.sdss.org/. The SDSS is managed by the Astrophysical
Research Consortium (ARC) for the Participating Institutions.  The
Participating Institutions include The University of Chicago, Fermilab, the
Institute for Advanced Study, the Japan Participation Group, The Johns Hopkins
University, Los Alamos National Laboratory, the Max-Planck-Institute for
Astronomy (MPIA), the Max-Planck-Institute for Astrophysics (MPA), New Mexico
State University, University of Pittsburgh, Princeton University, the United
States Naval Observatory, and the University of Washington.

\appendix

\section{Mass-to-light ratios of stellar populations}
\label{sec:ml-ratios-ssp}

Stellar M/L ratios of galaxies are generally based on evolutionary
population synthesis with the basic ingredient being simple stellar
populations (SSPs). This section outlines some issues related to M/L
determination: definition, IMF, metallicity, age, bursts, dust,
synthesis code. [See also \citet{PST04,KG07}.]

The following are common assumptions for estimating stellar masses.
The IMF is valid from 0.1 to 100 (or 120) $\Msun$.  Substellar
objects, $<0.1\Msun$, are {\em not} included; from the
\citet{Chabrier03} IMF, these add up to 5--10\% of the stellar mass.
Stellar remnants (white dwarfs, neutron stars, black holes) are
included; this is typically a 10--20\% factor in the \pegase\
models. Stellar mass is the remaining mass in stars and remnants as
opposed to the integral of the SF rate, i.e, $\mass =
(1-R) \int {\rm SFR}$ where $R$ is the recycled fraction.  This is the
definition used here and is appropriate considering the analysis of
\S~\ref{sec:mzr}.

\begin{figure*}
  \includegraphics[width=\middlecolsize\textwidth]{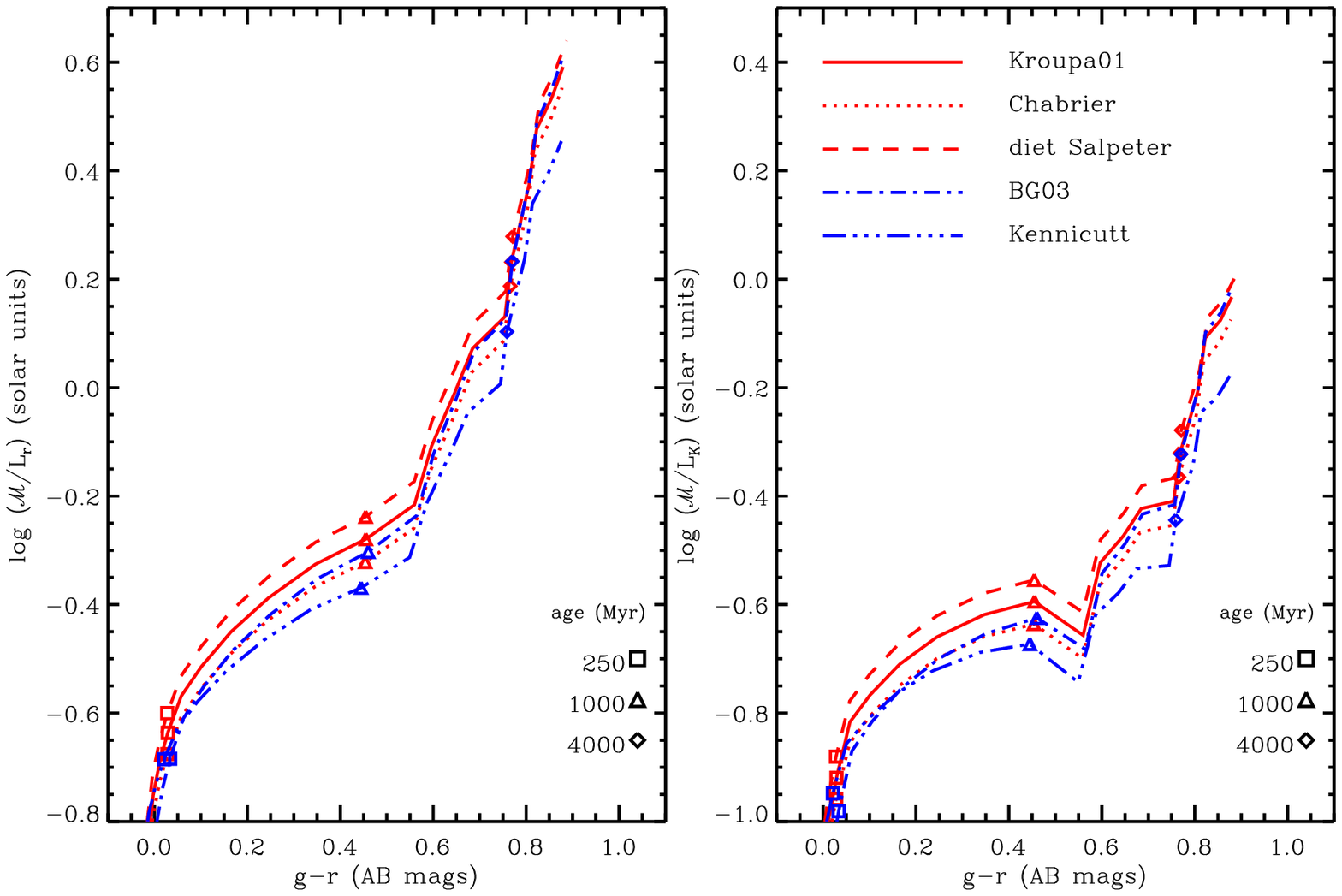}
\caption{M-L ratios of simple stellar populations computed using
  PEGASE with ages from 100\,Myr to 12\,Gyr versus colour at solar
  metallicity.  The left panel shows $r$-band M-L ratios while the
  right panel shows $K$-band M-L ratios. The tracks represent
  different IMFs \citep{Kroupa01,Chabrier03,BdJ01,BG03,Kennicutt83}.
  The upper-mass slopes of the IMFs are $\Gamma=1.3$, $1.3$, $1.35$,
  $1.15$, $1.5$, respectively (${\rm d}n/{\rm d}\log m \propto
  m^{-\Gamma}$).  The low-mass slopes are reduced at $m<0.5$ or
  $<1\Msun$ except for the diet Salpeter, which uses a 0.7 correction
  factor.}
\label{fig:ml-models-imfvar}
\end{figure*}

A significant factor is the choice of IMF.
Figure~\ref{fig:ml-models-imfvar} shows how M-L ratios in the $r$- and
$K$-bands vary with IMF choice.  The diet Salpeter has the highest
masses considered here as it was calibrated to have maximum M-L ratios
consistent with dynamical mass constraints \citep{BdJ01}.  Thus on the
assumption of a universally-applicable IMF, this is the most-massive
IMF at a given SSP colour that is valid. At the high-mass end of the
IMF, steep slopes implying fewer high-mass stars are ruled out by
measurements of cosmic luminosity densities \citep{BG03}.  Thus
reasonable M-L ratios are obtained with IMFs of Kroupa, Chabrier
(0.91--0.93), diet Salpeter (1.07--1.11), Baldry \& Glazebrook
(0.90--1.04), and Kennicutt (0.74--0.88), where the ranges in
parentheses are derived stellar masses relative to the Kroupa IMF at
SSP colours evaluated over the range of 0.2--0.85 in $g-r$. [If the
IMF varied with galaxy luminosity \citep{HG08}, over time
\citep{Dave08,vanDokkum08} and/or in star bursts \citep{fardal07},
this would clearly complicate the determination of stellar masses.]

\begin{figure*}
  \includegraphics[width=\doublecolsize\textwidth]{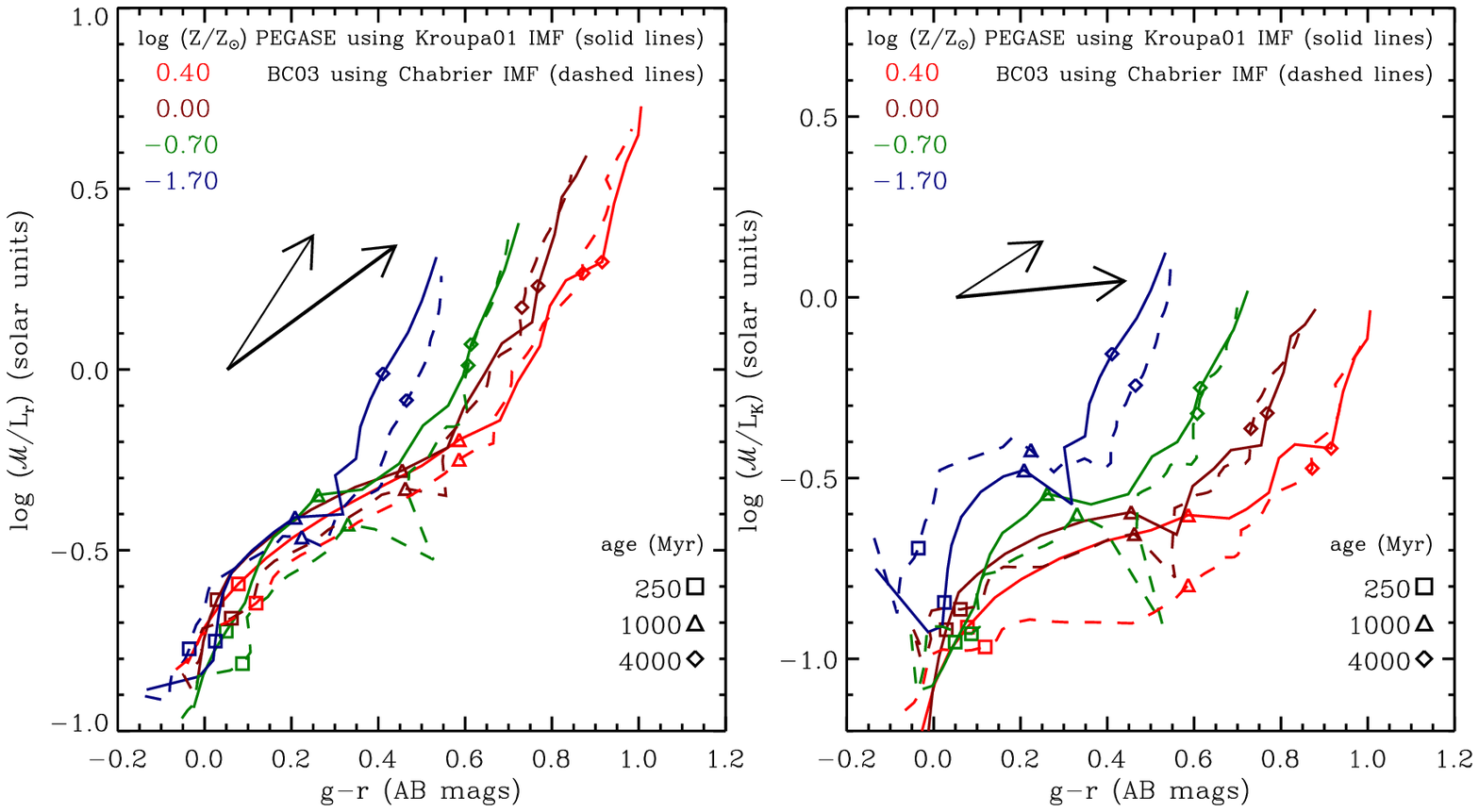}
\caption{M-L ratios of simple stellar populations with ages from
  100\,Myr to 12\,Gyr versus colour.  The left panel shows $r$-band
  M-L ratios while the right panel shows $K$-band M-L ratios. The
  tracks represent different metallicities colour-coded according to
  the legend in the top-left of each panel. The arrows correspond to
  the effect of $A_v=1{\rm\,mag}$ of dust attenuation for an SMC
  screen law (thicker line) and a $\lambda^{-0.7}$ power law.}
\label{fig:ml-models}
\end{figure*}

\begin{figure*}
  \includegraphics[width=\doublecolsize\textwidth]{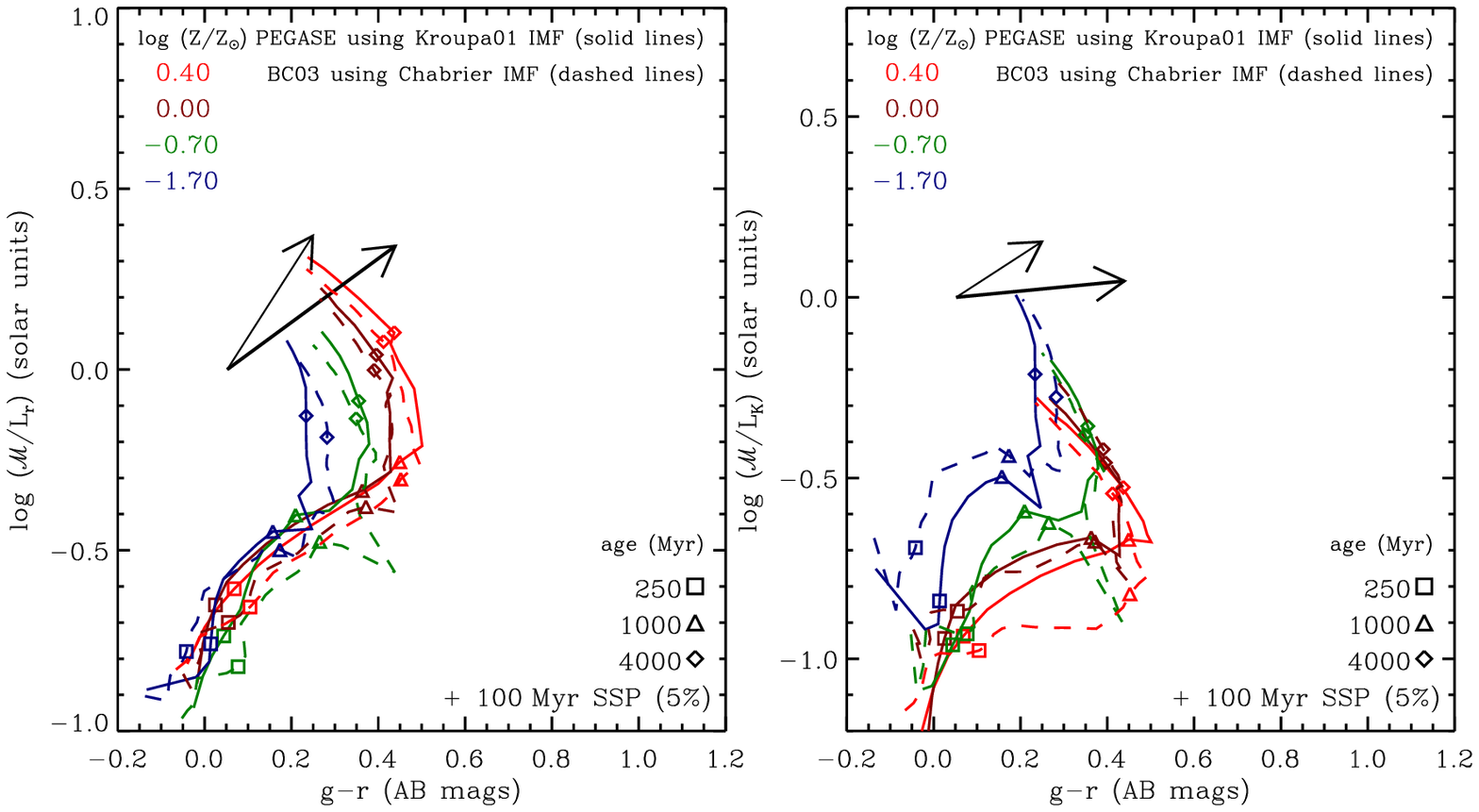}
\caption{As per Fig.~\ref{fig:ml-models} except a `burst'
  corresponding to an SSP of 100\,Myr contributing 5\% of the stellar
  mass has been added to each population. Since the SSPs are plotted
  from 100\,Myr, the tracks start in the same position as
  Fig.~\ref{fig:ml-models}.}
\label{fig:ml-models-withburst}
\end{figure*}

\begin{figure*}
  \includegraphics[width=\doublecolsize\textwidth]{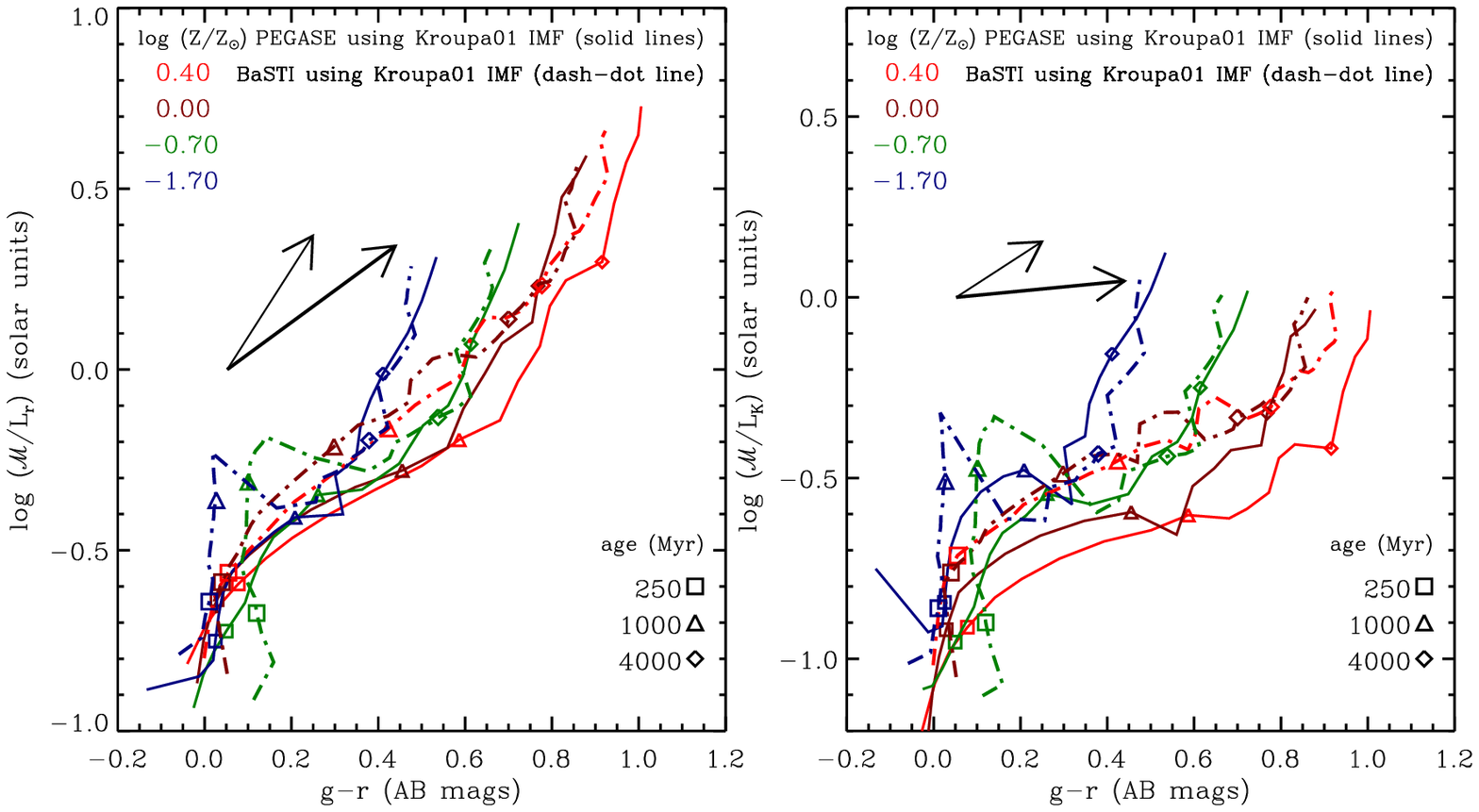}
\caption{As per Fig.~\ref{fig:ml-models} except comparing M-L
  ratios from BaSTI with PEGASE. The BaSTI models cover a larger
  metallicity range than implied by the legend with [Fe/H] from 0.4 to
  $-2.3$.}
\label{fig:ml-models-basti}
\end{figure*}

\begin{figure*}
  \includegraphics[width=\doublecolsize\textwidth]{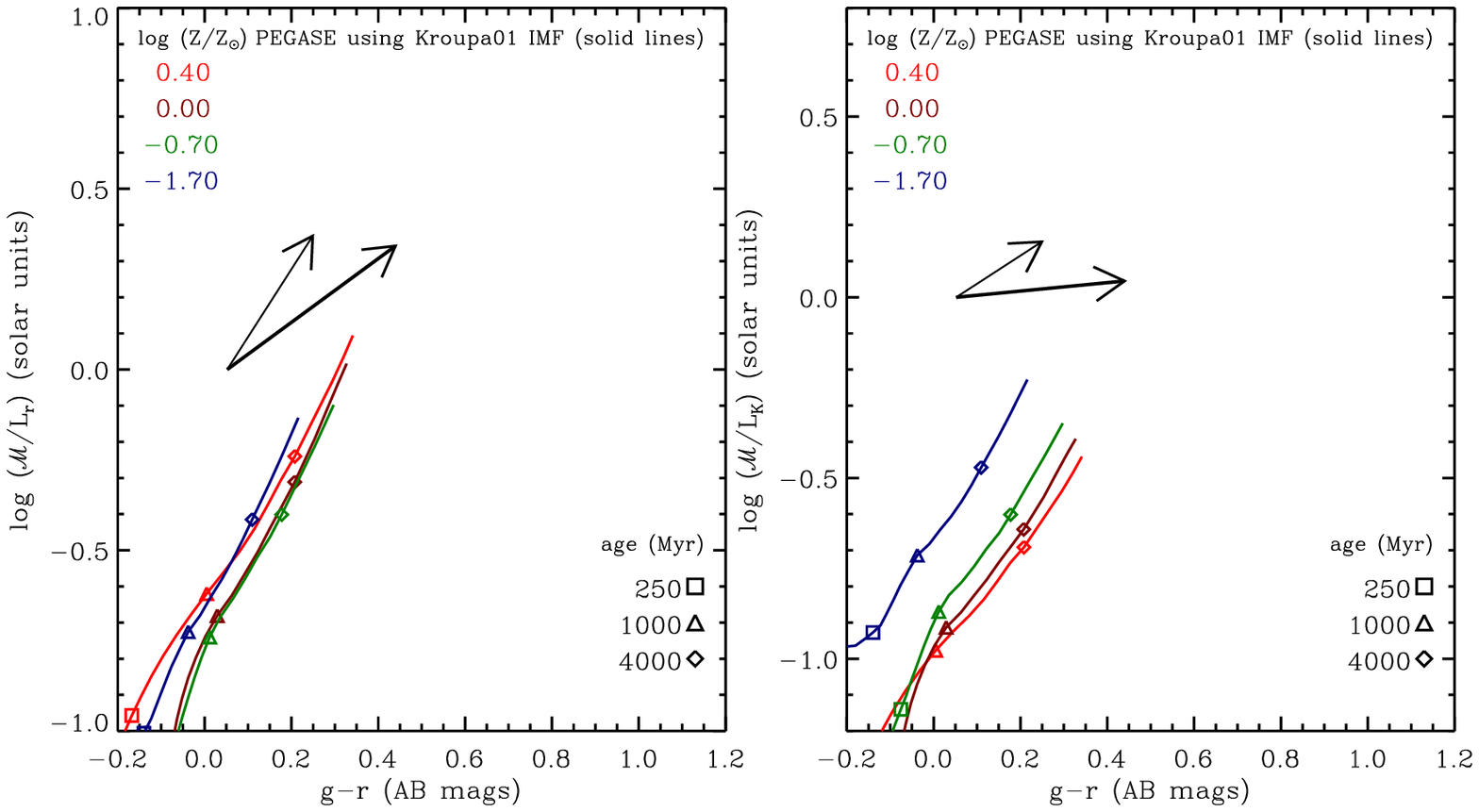}
\caption{As per Fig.~\ref{fig:ml-models} except for constant-rate
  SF histories.}
\label{fig:ml-models-cont}
\end{figure*}

A more significant consideration is the choice of prior (allowed SF
histories, etc.) and population synthesis code. The following figures
are provided to illuminate some of these considerations.
Figure~\ref{fig:ml-models} shows M-L ratio versus colour for SSPs over
a range of metallicities derived from \pegase\ \citep{FR97,FR99} and
\citet{BC03} models.  Figure~\ref{fig:ml-models-withburst} shows the
effect of adding a 100-Myr burst contributing 5\% of the stellar mass.
Figure~\ref{fig:ml-models-basti} shows a comparison between \pegase\ and
preliminary M-L ratios derived from \basti\ (M.~Salaris, private
communication; \citealt{pietrinferni04}).
Figure~\ref{fig:ml-models-cont} shows \pegase\ models for SF histories
with constant rate.

Last but not least, there is the complication of dust attenuation.
The arrows in Figs.~\ref{fig:ml-models}--\ref{fig:ml-models-cont} show
the effect of 1~magnitude of attenuation, at the $V$ band, on the
colours and M-L ratios: both for a Small Magellanic Cloud (SMC) screen
law \citep{Pei92} and a $\lambda^{-0.7}$ power law \citep{CF00}.  The
latter law allows for a range of attenuation to different parts of a
galaxy making it `greyer' than a screen law.

An example of the importance of the prior and dust law is demonstrated
by the fitting to the photometry of the NYU-VAGC sample
(\S~\ref{sec:nyu-vagc}).  Fitting only to photometry, as opposed to
spectral features, has the advantage of being less sensitive to
aperture bias but the disadvantage that colours are highly sensitive
to dust attenuation. Various fits including different dust laws were
tested (also fitting to $ugriz\,+\,JHK$).  It was found that using a
screen law significantly lowered the mass of luminous red galaxies
(early types), which were fitted with significant dust attenuation,
compared to assuming no dust.  The lower masses occurred because
younger-age models with dust reddening to reproduce the observed
colours give lower M-L ratios than older-age models with no dust,
i.e., the screen-law dust vector (change in M-L ratio as a function of
colour) is shallower than the age vector for red galaxies
(Fig.~\ref{fig:ml-models}).  Using a $\lambda^{-0.7}$ dust law did not
lower the masses but early-type galaxies were still fitted with
significant dust, which is not consistent with our general knowledge
of these galaxies, unless the fitting was restricted to solar
metallicity only. Comparable GSMFs were obtained either by allowing
for varying dust attenuation and assuming solar metallicity or by
allowing for a range of metallicities and assuming no dust. The former
is more appropriate for the high-mass galaxies while the latter is
more appropriate for low-mass, and low-metallicity, galaxies.

%%\bibliographystyle{mn}
%%\bibliography{galaxies,general,stars,surveys,two-d-f,fioc-rv}

\bsp

\label{lastpage}

\end{document}